\documentstyle[12pt]{article}
\begin{document}

\begin{flushright}
PITHA 94/25\\
hep-ph/9408316\\
July, 1994
\end{flushright}
\vspace*{2cm}
\LARGE\centerline{Energy correlation and asymmetry of}
\centerline{secondary leptons originating in}
\centerline{$H\to t\bar t$ and $H\to W^+W^-$}
\vspace*{1cm}
\large\centerline{T. Arens, U.D.J. Gieseler and L.M. Sehgal}
\centerline{III. Physikalisches Institut (A), RWTH Aachen,}
\centerline{D-52074 Aachen, Germany}
\normalsize

\begin{abstract}
We study the energy correlation of charged leptons produced in the 
decay of a heavy Higgs particle $H\to t\bar t\to bl^+\nu_l\bar 
bl^-\bar{\nu}_l$ and $H\to W^+W^-\to l^+\nu_ll^-\bar{\nu}_l.$ The possible 
influence of $CP$--violation in the $Ht\bar t$ and $HW^+W^-$ vertices 
on the energy spectrum of the secondary leptons is analyzed. The energy 
distribution of the charged leptons in the decay $H\to W^+W^-\to 
l^+\nu_ll^-\bar{\nu}_l$ is sensitive to the $CP$--parity of the 
Higgs particle and yields a simple criterion for distinguishing scalar 
Higgs from pseudoscalar Higgs.
\end{abstract}

\newpage
\section{Introduction}
We wish to report results on the energy spectrum and energy correlation of 
charged leptons produced in the reactions
\begin{eqnarray}
H&\to & t\bar t\to bl^+\nu_l\bar bl^-\bar{\nu}_l,\\
H&\to & W^+W^-\to l^+\nu_ll^-\bar{\nu}_l.
\end{eqnarray}
The above decays represent interesting leptonic signals of a heavy Higgs 
particle, that can be used to test the structure of Higgs couplings to 
fermions and gauge bosons \cite{nelson}. (Note that the reaction (2), in the 
standard 
model, is about 27 times more frequent than the ``gold--plated'' reaction 
$H\to ZZ\to \mu^+\mu^-\mu^+\mu^-$). We carry out the analysis in a 
general framework in which the couplings of the $H$ to $t\bar t$ and to 
$W^+W^-$ are given by:
\begin{eqnarray}
Ht\bar t\quad &:&\quad i(a+ib\gamma_5),\\
HW^+W^-\quad &:&\quad i2m_W^2\sqrt{G_F\sqrt{2}}(Bg_{\mu\nu}+\frac{D}{m_W^2}
\varepsilon_{\mu\nu\rho\sigma}p_{W^+}^{\rho}p_{W^-}^{\sigma}).
\end{eqnarray} 
Here $p_{W^+}$ and $p_{W^-}$ are the 4--momenta of the $W$--bosons. 
The terms proportional to $b$ and $D$ may arise as primary or induced 
effects in a generalized Higgs framework. Simultaneous presence of $a$ 
and $b$ or $B$ and $D$ is $CP$--violating \cite{peskin}. 
Results will be obtained   for the energy correlation of the two 
charged leptons in the $H$ rest frame. A special result is a simple 
criterion for distinguishing a scalar Higgs from a pseudoscalar Higgs 
particle on the basis of the energy spectrum of any single charged lepton in 
$H\to W^+W^-\to l^+\nu_ll^-\bar{\nu}_l$.

\section{$H\to t\bar t$}
The vertex $Ht\bar t$ (Eq. (3)) gives rise to the 
following differential decay rate for $H(P)\to t(p_t,s_+)\bar t(
p_{\bar t},s_-)$:
\begin{eqnarray}
\frac{\mbox{d}\Gamma}{\mbox{d}\Omega_t}(s_+,s_-)&=&\frac{\beta_t}{64\pi^2m_H}
\Bigl\{(|a|^2+|b|^2)(\frac{m^2_H}{2}-m_t^2+m_t^2s_+s_-)\nonumber\\
&&+(|a|^2-|b|^2)(Ps_+Ps_--\frac{m_H^2}{2}s_+s_-+m_t^2s_+s_--m_t^2)
\nonumber\\
&&-\mbox{Re}(ab^*)\varepsilon (P,Q,s_+,s_-)-2\mbox{Im}(ab^*)m_tP(s_++s_-)
\Bigr\}, 
\end{eqnarray}
where $P\equiv p_t+p_{\bar t}$, $Q=p_t-p_{\bar t}$, and  
$s_+$ and $s_-$ denote the polarization vectors of $t$ and $\bar t$, 
respectively. $\beta_t=\sqrt{1-4m_t^2/m_H^2}$ is the velocity of  the top 
quarks in the Higgs rest frame. The symbol $\varepsilon (a,b,c,d)$ means 
$\varepsilon_{\mu\nu\rho\sigma}a^{\mu}b^{\nu}c^{\rho}d^{\sigma}$ with 
$\varepsilon_{0123}=+1$. The terms proportional to $\mbox{Re}(ab^*)$ and 
$\mbox{Im}(ab^*)$ represent the $CP$--violating part of the differential 
decay rate.\\
\indent Using the method of Kawasaki, Shirafuji and Tsai \cite{tsai}, the 
differential decay rate $\displaystyle\frac{\mbox{d}\Gamma}{\mbox{d}\Omega_t}
(s_+,s_-)$ 
yields the following normalized energy correlation of the charged leptons 
produced in the decay $H\to t\bar t\to l^+l^-+\cdots $ [F 1]:
\begin{eqnarray}
\frac{1}{\Gamma}\frac{\mbox{d}\Gamma}{\mbox{d}x\mbox{d}x'}(H\to 
t\bar t\to l^+l^-+\cdots )&=& 
f(x)f(x')-\frac{1}{\beta_t^2}g(x)g(x')\nonumber\\
&&+\frac{2\mbox{Im}(ab^*)}{|a|^2\beta_t^2+|b|^2}\Bigl [f(x')g(x)-f(x)g(x')
\Bigr ],
\end{eqnarray}
where $x$ and $x'$ are the reduced energies
\begin{eqnarray}
x=\frac{2E(l^+)}{m_t}\sqrt{\frac{1-\beta_t}{1+\beta_t}}\qquad ,\qquad
x'=\frac{2E(l^-)}{m_t}\sqrt{\frac{1-\beta_t}{1+\beta_t}},
\end{eqnarray}
$E(l^+)$ and $E(l^-)$ being the energies of the final leptons $l^+$  and $l^-$ 
in the Higgs rest frame. $x$ and $x'$ are bounded by
\begin{eqnarray}
\frac{m_W^2}{m^2_t}\frac{1-\beta_t}{1+\beta_t}\le x,x'\le 1,
\end{eqnarray}
assuming the narrow width approximation for the $W$--bosons in the top 
decay. The functions $f$ and $g$ are defined as follows (see \cite{arens}):
\newline
\newline
1. $\displaystyle\frac{m_W^2}{m_t^2}\ge\displaystyle\frac{1-\beta_t}{1+
\beta_t}$
\begin{eqnarray*}
f(x)=\frac{3}{2W}\frac{1+\beta_t}{\beta_t}&&\!\!\!\!\!\!\!\!
\left\{ \begin{array}{l@{\qquad}l}
-2\displaystyle\frac{m_W^2}{m_t^2}+\displaystyle\frac{m_W^4}{m_t^4}+2x
\displaystyle\frac{1+\beta_t}{1-\beta_t}-x^2(\displaystyle\frac{1+\beta_t}
{1-\beta_t})^2 & :I_1\\
1-2\displaystyle\frac{m_W^2}{m_t^2}+\displaystyle\frac{m_W^4}
{m_t^4} & :I_2\\
1-2x+x^2 & :I_3\\
\end{array}\right.\\
g(x)=\frac{3}{W}\frac{(1+\beta_t)^2}{\beta_t}&&\!\!\!\!\!\!\!\!
\left\{ \begin{array}{l@{\quad}l}
-x\displaystyle\frac{m_W^2}{m_t^2}+x^2\displaystyle\frac{1+\beta_t}{1-\beta_t}
+x\ln\displaystyle\frac{m_W^2}{m_t^2}-x\ln (x\displaystyle\frac{1+\beta_t}
{1-\beta_t})\\
+\displaystyle\frac{1/2}{1+\beta_t}\big [-2\displaystyle\frac{m_W^2}
{m_t^2}+\displaystyle\frac{m_W^4}{m_t^4}+2x\displaystyle\frac{1+\beta_t}
{1-\beta_t}-x^2(\displaystyle\frac{1+\beta_t}{1-\beta_t})^2\big ] & :I_1\\
\\
x-x\displaystyle\frac{m_W^2}{m_t^2}+x\ln\displaystyle\frac{m_W^2}{m_t^2}
+\displaystyle\frac{1/2}{1+\beta_t}\big [1-2\displaystyle\frac{m_W^2}{m_t^2}+
\displaystyle\frac{m_W^4}{m_t^4}\big ] & :I_2\\
x-x^2+x\ln x+\displaystyle\frac{1/2}{1+\beta_t}\big [1-2x+x^2\big ] & :I_3\\
\end{array}\right.
\end{eqnarray*}
where the intervals $I_i$ are given by:\newline
\begin{eqnarray*}
\begin{array}{c@{\quad : \quad}c@{\quad\le\quad}c@{\quad\le\quad}c}
I_1 & \displaystyle\frac{m_W^2}{m_t^2}\displaystyle\frac{1-\beta_t}{1+\beta_t}
&x&\displaystyle\frac{1-\beta_t}{1+\beta_t},\\
I_2 & \displaystyle\frac{1-\beta_t}{1+\beta_t}&x&\displaystyle\frac{m_W^2}
{m_t^2},\\
I_3 & \displaystyle\frac{m_W^2}{m_t^2}&x&1.
\end{array}
\end{eqnarray*}
2. $\displaystyle\frac{m_W^2}{m_t^2}\le\displaystyle \frac{1-\beta_t}{1
+\beta_t}$
\begin{eqnarray*}
f(x)=\frac{3}{2W}\frac{1+\beta_t}{\beta_t}&&\!\!\!\!\!\!\!\!
\left\{ \begin{array}{l@{\qquad}l}
-2\displaystyle\frac{m_W^2}{m_t^2}+\displaystyle\frac{m_W^4}{m_t^4}+2x
\displaystyle\frac{1+\beta_t}{1-\beta_t}-x^2(\displaystyle\frac{1+\beta_t}{1-
\beta_t})^2 & :I_4\\
-2x+x^2+2x\displaystyle\frac{1+\beta_t}{1-\beta_t}-x^2(\displaystyle\frac{1+
\beta_t}{1-\beta_t})^2 & :I_5\\
1-2x+x^2 & :I_6\\
\end{array}\right.\\
g(x)=\frac{3}{W}\frac{(1+\beta_t)^2}{\beta_t}&&\!\!\!\!\!\!\!\!
\left\{ \begin{array}{l@{\quad}l}
-x\displaystyle\frac{m_W^2}{m_t^2}+x^2\displaystyle\frac{1+\beta_t}{1-\beta_t}+
x\ln\displaystyle\frac{m_W^2}{m_t^2}-x\ln (x\displaystyle\frac{1+\beta_t}
{1-\beta_t})\\
+\displaystyle\frac{1/2}{1+\beta_t}\big [-2\displaystyle\frac{m_W^2}
{m_t^2}+\displaystyle\frac{m_W^4}{m_t^4}+2x\displaystyle\frac{1+\beta_t}
{1-\beta_t}-x^2(\displaystyle\frac{1+\beta_t}{1-\beta_t})^2\big ] & :I_4\\
\\
-x^2+x^2\displaystyle\frac{1+\beta_t}{1-\beta_t}+x\ln\displaystyle\frac{1-
\beta_t}{1+\beta_t}+\displaystyle\frac{1/2}{1+\beta_t}\big [-2x+x^2 \\
+2x\displaystyle\frac{1+\beta_t}{1-\beta_t}-x^2(\displaystyle\frac{1+\beta_t}
{1-\beta_t})^2\big ] & :I_5\\
\\
x-x^2+x\ln x+\displaystyle\frac{1/2}{1+\beta_t}\big [1-2x+x^2\big ] &:I_6\\
\end{array}\right.
\end{eqnarray*}
with the intervals $I_i$:\newline
\begin{eqnarray*}
\begin{array}{c@{\quad :\quad}c@{\quad\le\quad}c@{\quad\le\quad}c}
I_4&\displaystyle\frac{m_W^2}{m_t^2}\displaystyle\frac{1-\beta_t}{1+\beta_t}
&x&\displaystyle\frac{m_W^2}{m_t^2},\\
I_5&\displaystyle\frac{m_W^2}{m_t^2}&x&\displaystyle\frac{1-\beta_t}{1+\beta_t},\\
I_6&\displaystyle\frac{1-\beta_t}{1+\beta_t}&x&1,
\end{array}
\end{eqnarray*}
and 
\begin{eqnarray}
W=(1-\frac{m_W^2}{m_t^2})^2(1+2\frac{m_W^2}{m_t^2}).
\end{eqnarray}
The normalizations of $f$ and $g$ are
\begin{eqnarray}
\int f(x)dx=1,\nonumber\\
\int g(x)dx=0.
\end{eqnarray}
The functions $f$  and $g$ represent the spin--independent and spin--dependent 
parts of the lepton spectrum in $t$--decay. Eq. (6) can also be written as
\begin{eqnarray}
\frac{1}{\Gamma}\frac{\mbox{d}\Gamma}{\mbox{d}x\mbox{d}x'}(H\to t\bar t\to 
l^+l^-+\cdots ) = S_t(x,x')+\Delta A_t(x,x'),
\end{eqnarray}
where
\begin{eqnarray}
S_t(x,x')&=&f(x)f(x')-\frac{1}{\beta_t^2}g(x)g(x'),\nonumber\\
A_t(x,x')&=&f(x')g(x)-f(x)g(x'),\nonumber\\
\Delta &=& \frac{2\mbox{Im}(ab^*)}{|a|^2\beta^2_t+|b|^2}.
\end{eqnarray}
$S_t(x,x')$ and $A_t(x,x')$ represent the symmetric and antisymmetric part of 
the energy correlation. These are plotted in Figs. (1a) and (1b).\\
\indent The symmetric ($CP$--conserving) part of the two--dimensional 
distribution $\displaystyle\frac{1}{\Gamma}\frac{\mbox{d}\Gamma}{\mbox{d}x
\mbox{d}x'}$ does not depend on the coupling 
constants $a$ and $b$. This means that in the $CP$--conserving limit 
the energy correlation of secondary leptons arising from $H\to t\bar t$ is 
independent of the $CP$--parity of the decaying Higgs particle.\\
\indent Integration over $x$ or $x'$ yields the single lepton energy spectra
\begin{eqnarray}
\frac{1}{\Gamma}\frac{\mbox{d}\Gamma}{\mbox{d}x}(H\to t\bar t\to 
l^{\pm}+\cdots )= f(x)\pm \Delta g(x).
\end{eqnarray}
Eq. (13) agrees with the energy spectrum obtained by Chang and Keung 
\cite{chang} using a different method. The single energy spectra are 
plotted in Fig. 2. The parameter $\Delta$ is calculated within a 
2--Higgs Doublet Model in Refs. \cite{chang,grad}.

\section{$H\to W^+W^-$}
The differential decay rate for the reaction $\!H(P)\!\to\! W^+W^-\!\!\to \!
l^+(q_1)\nu_l(q_2)l^-(q_3)\bar{\nu}_l(q_4)$, arising from the $HW^+W^-$ vertex 
given in Eq. (4), is 
\begin{eqnarray}
\mbox{d}^{8}\Gamma=8\sqrt{2}\,\frac{G_{F}}{m_{H}}D_{W} \left[
              |B|^{2}\,{\cal S}+\frac{|D|^{2}}{m_{W}^{4}}\,{\cal P}
              +\frac{\mbox{Re}(BD^{*})}{m_{W}^{2}}\,{\cal Q}
              -\frac{\mbox{Im}(BD^{*})}{m_{W}^{2}}\,{\cal R} \right]
\cdot\mbox{d}Lips . 
\end{eqnarray}
The Lorentz invariant phase space is given by 
\begin{eqnarray}
\mbox{d}Lips = (2\pi )^4\delta^{(4)}(P-q_1-
q_2-q_3-q_4)\prod_{i=1}^4\frac{\mbox{d}^3q_i}{(2\pi )^32q_i^0}.
\end{eqnarray}
In the massless fermion approximation,
\begin{eqnarray}
{\cal S}&=&(q_2\cdot q_3)(q_1\cdot q_4),\nonumber\\
{\cal P}&=&-\Bigl \{(q_2\cdot q_3)(q_1\cdot q_4)-(q_2\cdot q_4)
(q_1\cdot q_3)\Bigr\}^2\nonumber\\
&&+\frac{m_W^4}{4}\Bigl\{ (q_2\cdot q_3)^2+(q_1\cdot q_4)^2+
2(q_2\cdot q_4)(q_1\cdot q_3)-\frac{m_W^4}{4}\Bigr\},\nonumber\\
{\cal Q}&=&\varepsilon (q_1,q_2,q_3,q_4)\Bigl\{ (q_2\cdot q_3)+
(q_1\cdot q_4)\Bigr\},\nonumber\\
{\cal R}&=&\Bigl\{ (q_2\cdot q_3)-(q_1\cdot q_4)\Bigr\}\Bigl\{
\frac{m_W^4}{4}+(q_2\cdot q_3)(q_1\cdot q_4)-
(q_2\cdot q_4)(q_1\cdot q_3)\Bigr\},
\end{eqnarray}
while $D_W$ is the propagator factor
\begin{eqnarray}
D_W=m_W^4\prod^{2}_{j=1}
                \frac{g^2}{(s_j-m_W^2)^2+m_W^2\Gamma_W^2},
\end{eqnarray}
with $s_1=(q_1+q_2)^2$, $s_2=(q_3+q_4)^2$. In  the narrow width 
approximation, the total decay rate is given by
\begin{eqnarray}
\Gamma (H\to W^+W^-\to l^+\nu_ll^-\bar{\nu}_l)&=& 
       \frac{g^6m_H^3\beta_W}{9\cdot 2^{16}\pi^3\Gamma^2_W}\Big\{\nonumber\\
&&|B|^{2}(3-2\beta_W^2+3\beta_W^4)+8|D|^{2}\beta_W^2 \Big\},
\end{eqnarray} 
in agreement with the result of Osland and Skjold \cite{osland}.\\
\indent We now introduce scaled energy variables in the $H$ rest frame:
\begin{eqnarray}
y = \frac{4E(l^+)}{m_H} \qquad , \qquad y' = \frac{4E(l^-)}{m_H},
\end{eqnarray}
which are bounded by
\begin{eqnarray}
1-\beta_W\le y,y'\le 1+\beta_W ,
\end{eqnarray}
where $\beta_W=\sqrt{1-4m_W^2/m_H^2}$. The two--dimensional spectrum in the 
variables $y$ and $y'$ is then given by
\begin{eqnarray}
\frac{1}{\Gamma}\frac{\mbox{d}\Gamma}{\mbox{d}y\mbox{d}y'}
(H\!\!&\!\!\to\!\!&\!\! W^+W^-\to l^+l^-+\cdots )
=\frac{1}{|B|^{2}(3-2\beta_W^{2}+3\beta_W^4)
+8|D|^{2}\beta_W^{2}} \cdot \frac{9}{32\beta_W^6} \cdot \nonumber\\
\!\!&\!\!\!\!&\!\! \bigg\{|B|^2\Big[(3+2\beta_W^{2}
+3\beta_W^{4})((y-1)^{2}-\beta_W^{2})((y'-1)^{2}-\beta_W^{2})\nonumber\\
&& \hspace{5.65cm}           +2\beta_W^{2}(1-\beta_W^{2})^{2}(y-y')^2\Big]
\nonumber\\
\!\!&\!\!\!\!&\!\! +4\beta_W^2|D|^{2} \Big[ ((y-1)^{2}+\beta_W^{2})
((y'-1)^{2}+\beta_W^2)-4\beta_W^2(y-1)(y'-1)\Big]\nonumber\\
\!\!&\!\!\!\!&\!\! +8\beta_W^2\mbox{Im}(BD^*) (1-\beta_W^2)
\Big[ \beta_W^2-(y-1)(y'-1)\Big](y-y') \bigg\}.
\end{eqnarray}
Neglecting terms proportional to $|D|^2$, the correlation can be written as
\begin{eqnarray}
\frac{1}{\Gamma}\frac{\mbox{d}\Gamma}{\mbox{d}y\mbox{d}y'}(H\to W^+W^-\to 
l^+l^-+\cdots )
=S_W(y,y')&+&\frac{\mbox{Im}(BD^*)}{|B|^2}A_W(y,y')\nonumber\\
&+&O(|D|^2/|B|^2).
\end{eqnarray}
Here $S_W(y,y')$ and $A_W(y,y')$ represent the symmetric and antisymmetric 
parts of the energy correlation of the charged leptons, the latter being 
multiplied by the $CP$--violating coefficient $\mbox{Im}(BD^*)/|B|^2$. These 
functions are plotted in Figs. (3a) and (3b).\\
\indent There is an interesting difference in the energy characteristics of the 
leptons emanating from $H\to W^+W^-\to l^+\nu_ll^-\bar{\nu}_l$, dependent 
on whether $H$ is a scalar $(0^+)$ or pseudoscalar $(0^-)$ particle. The 
correlated energy spectrum of the $l^+l^-$ pair can be derived from Eq. (21) 
by taking $D=0$ (scalar case) or $B=0$ (pseudoscalar case), with the 
result
\begin{eqnarray}
\frac{1}{\Gamma}\frac{\mbox{d}\Gamma (0^+)}{\mbox{d}y\mbox{d}y'}=S_W(y,y')
\!\!&\!\!=\!\!&\!\!\frac{9}{32\beta_W^6}\frac{1}{3-2\beta_W^2+3\beta_W^4}
\Big[2\beta_W^{2}(1-\beta_W^{2})^{2}(y-y')^2
\nonumber\\
\!\!\!\!&\!\!\!\!&\!\!\!+(3+2\beta_W^{2}+3\beta_W^{4})((y-1)^{2}-
\beta_W^{2})((y'-1)^{2}-\beta_W^{2})\Big] , \\
\frac{1}{\Gamma}\frac{\mbox{d}\Gamma (0^-)}{\mbox{d}y\mbox{d}y'}=P_W(y,y')
\!\!&\!\!=\!\!&\!\!\frac{9}
{64\beta_W^6}\Big[ ((y-1)^{2}+\beta_W^{2})((y'-1)^{2}+\beta_W^{2})\nonumber\\
\!\!\!\!&\!\!\!\!&\!\!\!\!-4\beta_W^{2}(y-1)(y'-1)\Big].
\end{eqnarray}
These two functions are strikingly different, as shown in Figs. (3a) and (3c). 
This difference persists even if we consider the energy spectrum of a single 
lepton. Integrating Eqs. (23, 24) over $y'$, we have
\begin{eqnarray}
\frac{1}{\Gamma}\frac{\mbox{d}\Gamma (0^+)}{\mbox{d}y}&=&\frac{3}{2\beta_W}
\frac{1+\beta_W^4-2(y-1)^2}{3-2\beta_W^2+3\beta_W^4},\\
\frac{1}{\Gamma}\frac{\mbox{d}\Gamma (0^-)}{\mbox{d}y}&=&\frac{3}{8\beta_W^3}
(\beta_W^2+(y-1)^2).
\end{eqnarray}
These distributions are clearly quite distinct (Fig. 4) and provide a 
simple criterion for distinguishing $0^+$ and $0^-$ objects decaying into 
$W^+W^-$ pairs. Indeed, the single lepton spectra (Eqs. (25), (26)) are also 
valid for the inclusive process $H\to W^+W^-\to l^{\pm}X$, where only one of 
the $W$--bosons decays leptonically. The difference in the lepton energy 
spectrum for the $0^+$  and $0^-$ cases is intimately related to the 
different helicity structure of the $W$--bosons produced in the two cases 
\cite{zerwas}. It should be stressed that the correlations and spectra 
given above (Eqs. (21)--(26)) refer directly to energies measured   in the $H$ 
rest frame, and do not require reconstruction of the decay planes of $W^+$ 
and $W^-$. In this respect, the present criterion provides a useful 
alternative to other criteria that have recently been proposed in the 
literature \cite{zerwas, barger}. Finally, we note that the energy spectrum 
in the $0^+$ case agrees with that obtained by Chang and Keung \cite{chang}, 
after correction of a minor typographical error [F 2].\newline\newline 

\indent One of us (T.A.) acknowledges a stipend from the NRW 
Graduierten\-f\"or\-derungs\-pro\-gramm. This research has been supported 
by the BMFT (German Ministry of Research and Technology).

\newpage
\Large
\noindent{\bf Footnotes}
\normalsize
\begin{itemize}
\item[[F 1]] Some of the essential steps in the procedure of 
Ref. \cite{tsai} can 
be found in Ref. \cite{arens}.
\item[[F 2]] Eq. (15) of Ref. \cite{chang} should read
\begin{eqnarray*}
\frac{1}{N}\frac{\mbox{d}N}{\mbox{d}x(l^{\pm})}=\Biggl (\frac{(1+\beta_W^2)^2}
{3-2\beta_W^2+3\beta_W^4}\Biggr )
\frac{3\bigl[\beta_W^2-(1-x)^2\bigr ]}{4\beta_W^3}+
\sum_{s=-1,+1}\cdots .
\end{eqnarray*}
\end{itemize}

\newpage\Large
\noindent{\bf Figure Captions}
\normalsize
\begin{itemize}
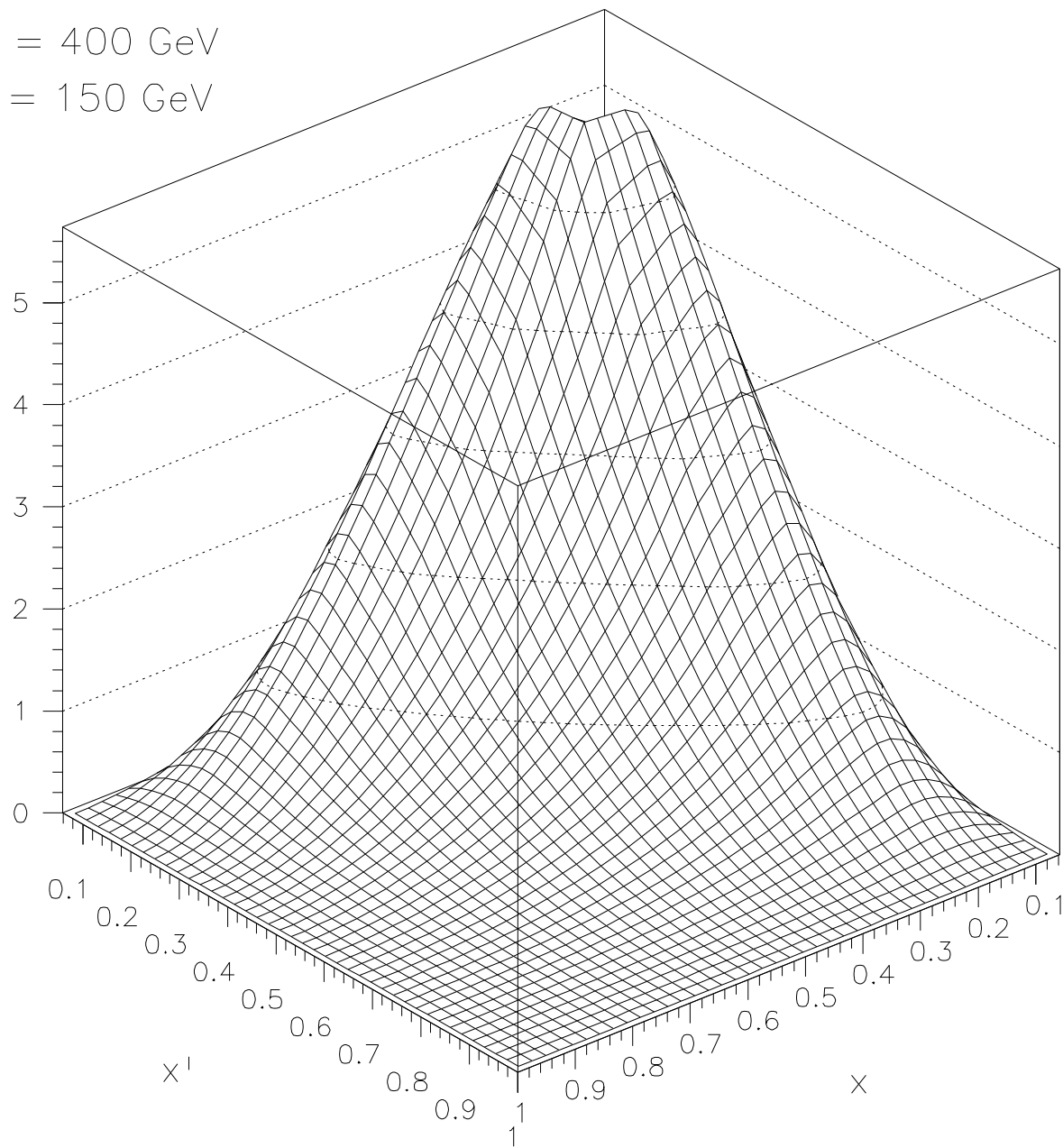
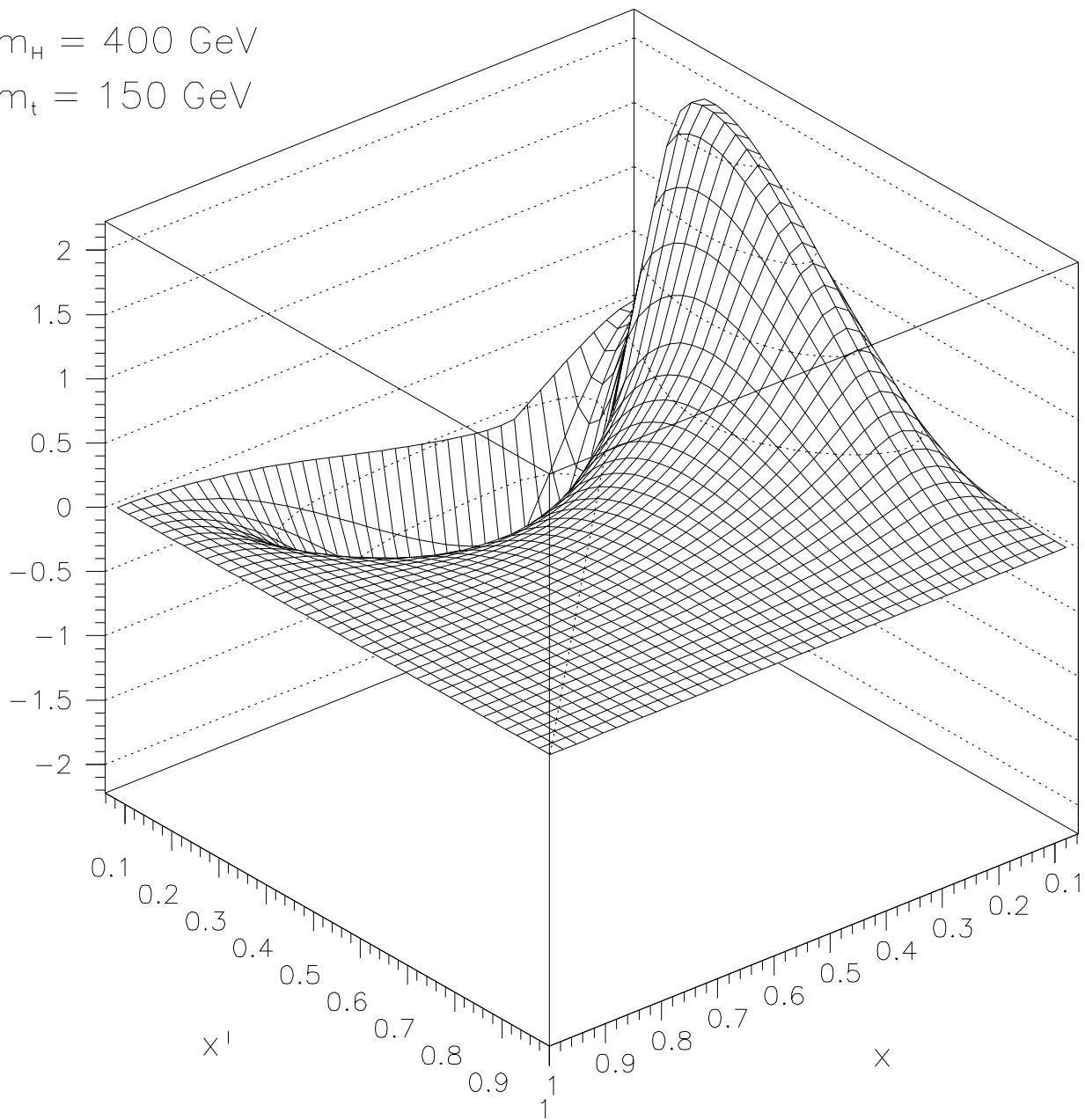
\item[Fig. 1.] $CP$--conserving (a) and $CP$--violating (b) part of the 
normalized 
energy correlation in the decay $H\to t\bar t\to l^+l^-+\cdots$ for $m_H=400$ 
GeV and $m_t=150$ GeV.
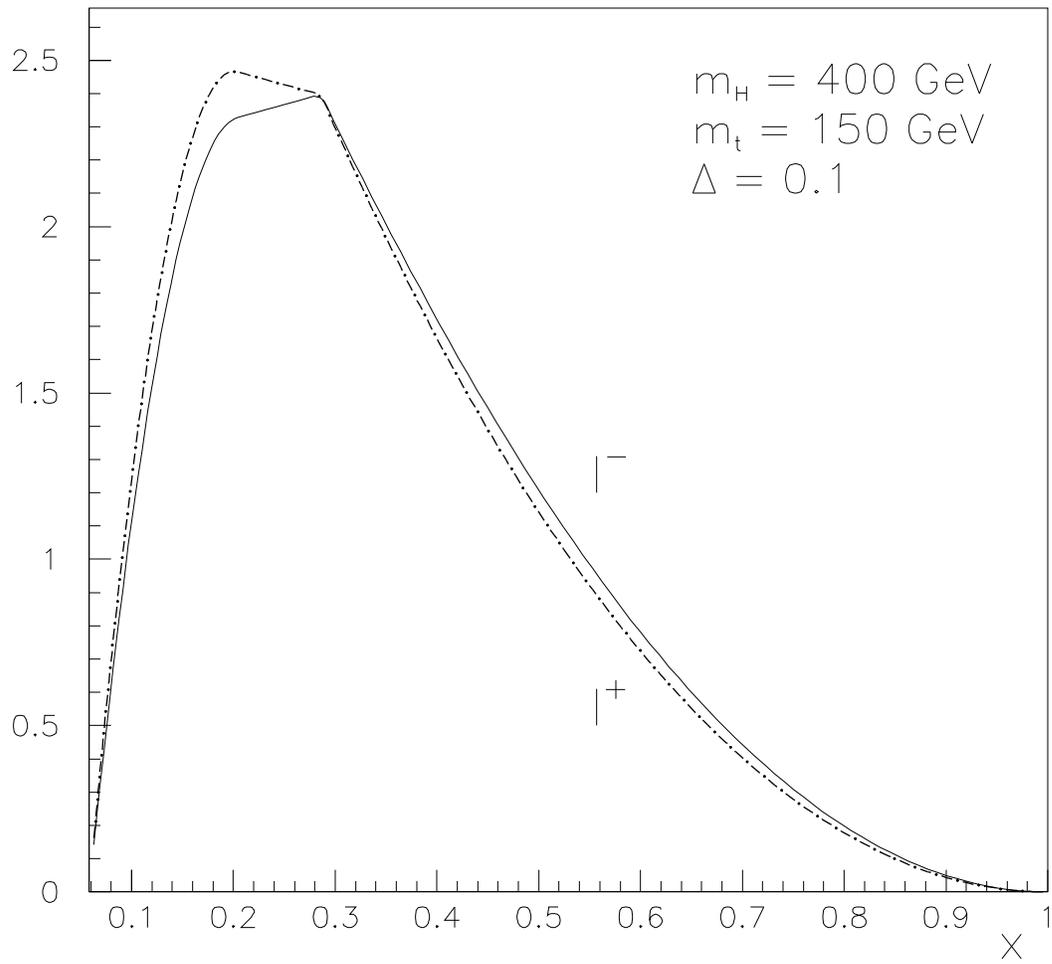
\item[Fig. 2.] Single particle energy spectra of $l^+$ (dotted curve) and $l^-$ 
(full curve) in the decay 
$H\to t\bar t$ for $\Delta = 0.1$, $m_H=400$ GeV and $m_t=150$ GeV.
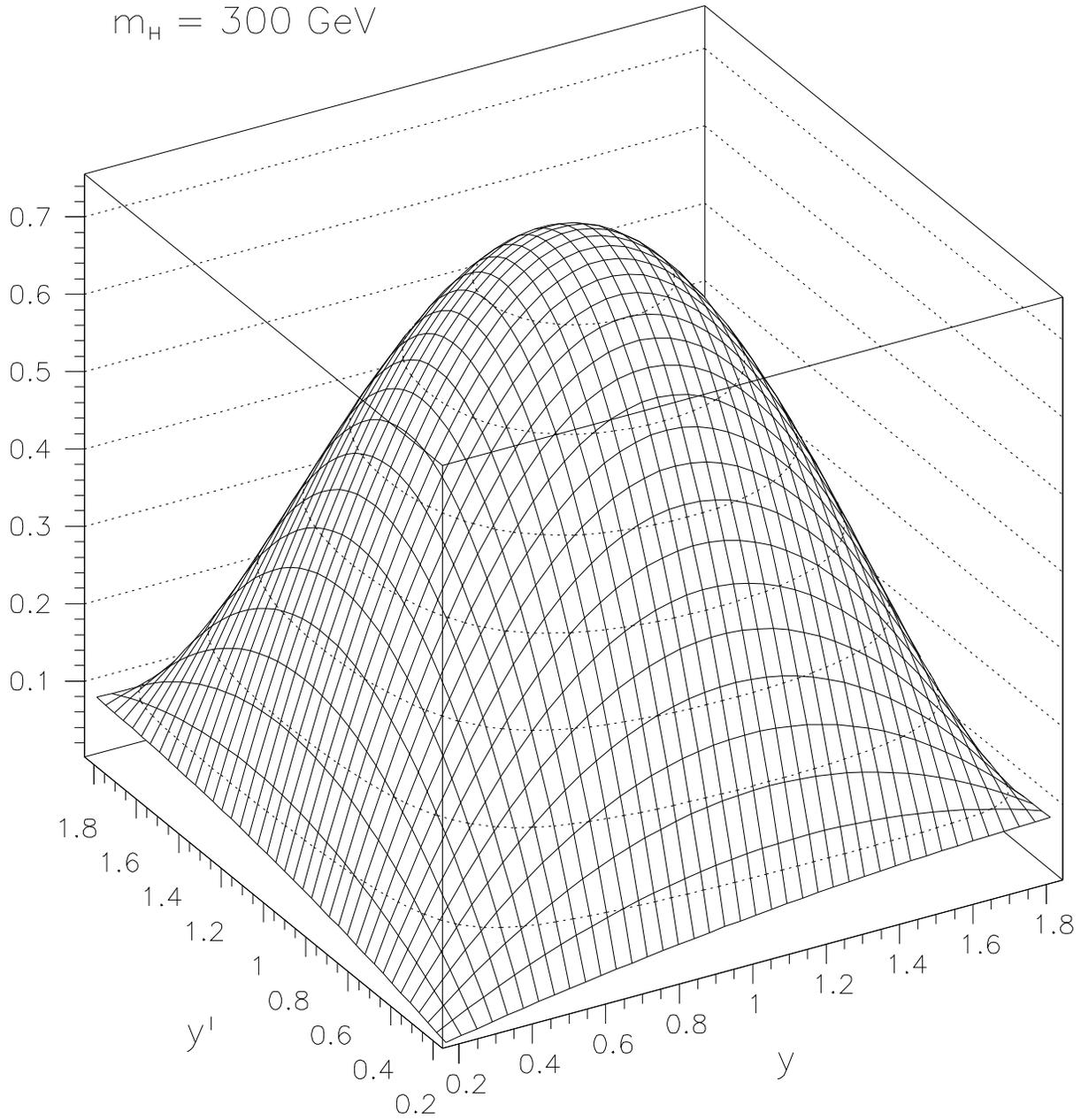
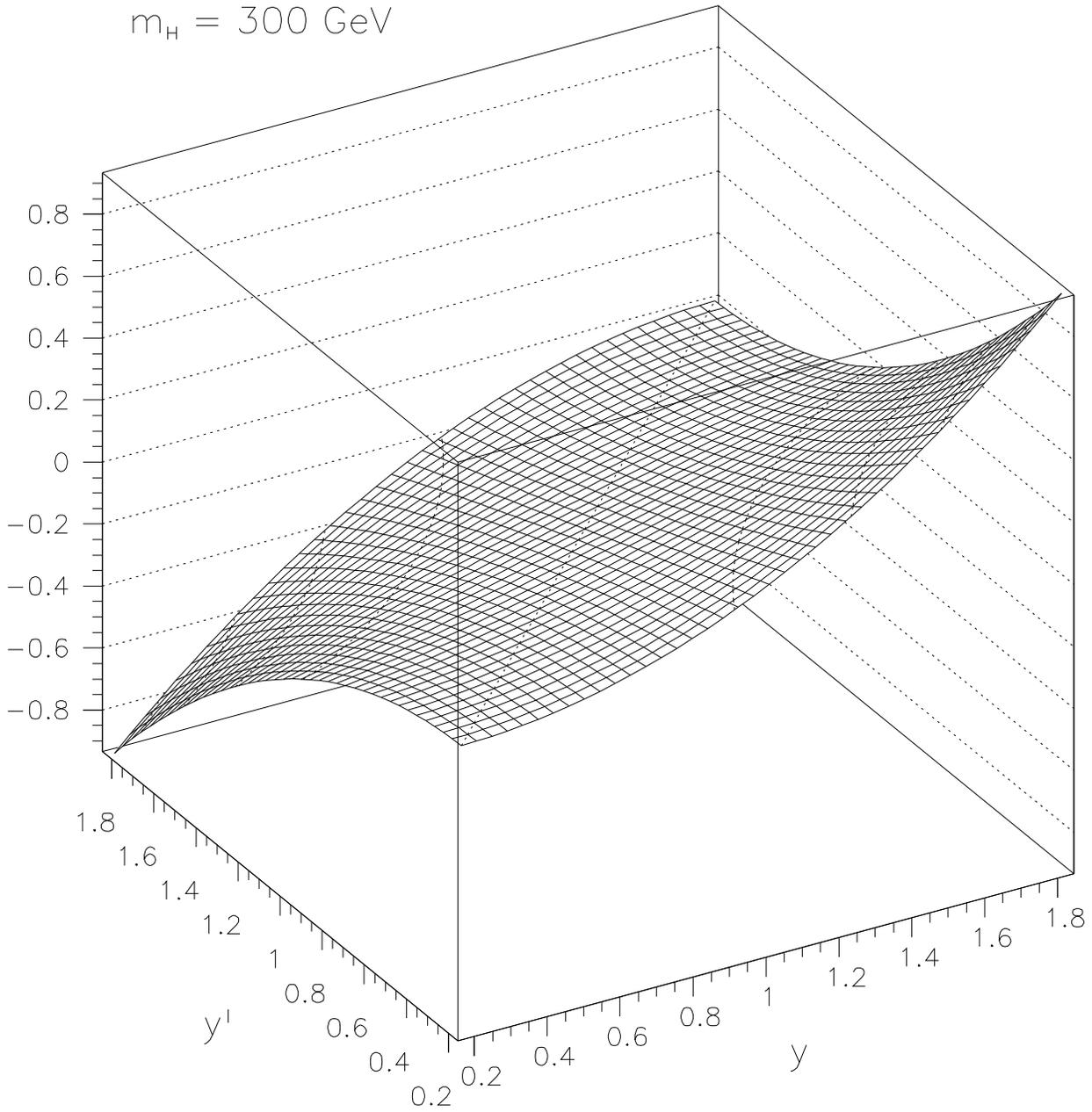
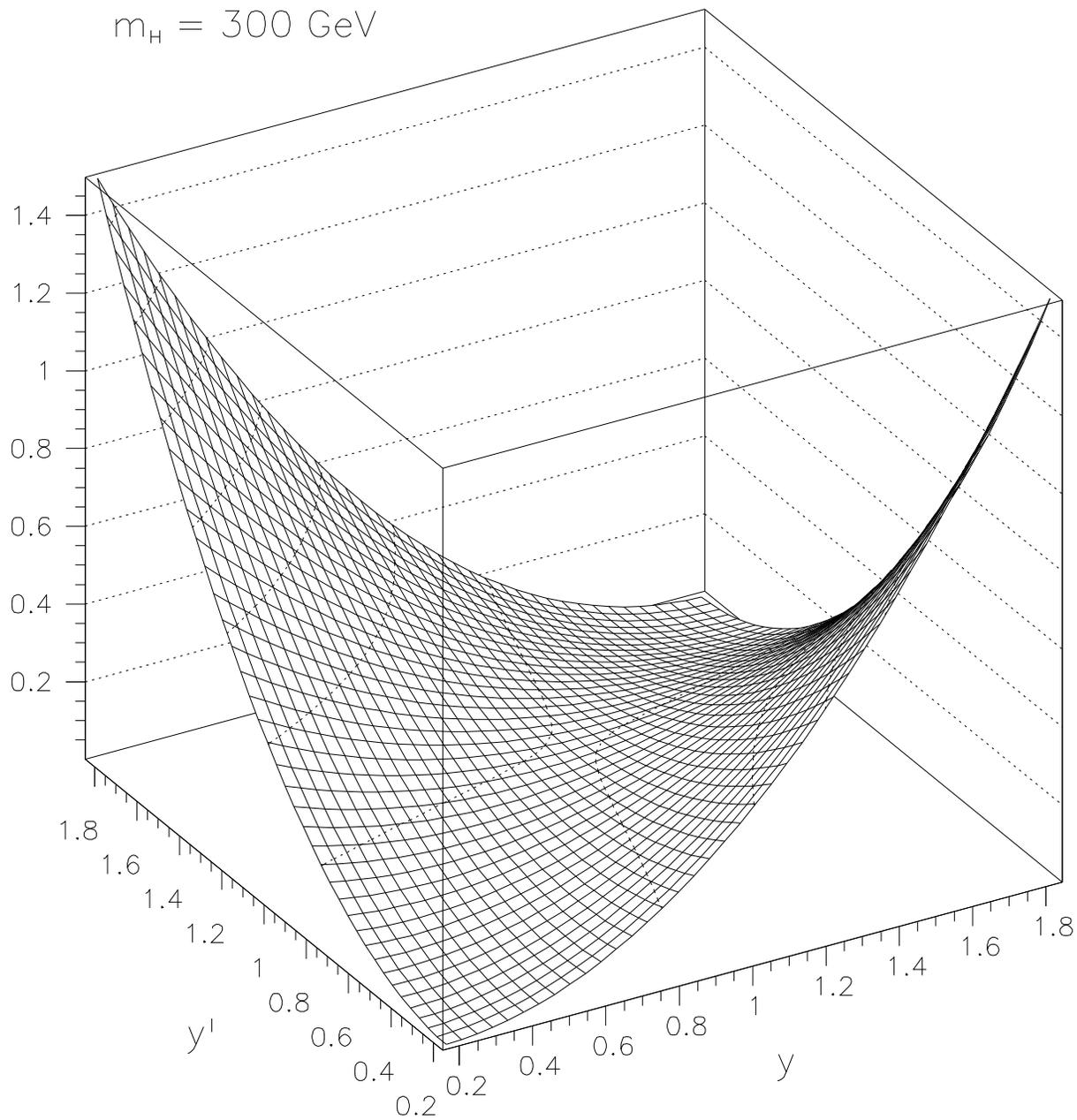
\item[Fig. 3.] $CP$--conserving (a) and $CP$--violating (b) part of the 
normalized 
energy correlation in the decay $H\to W^+W^-\to l^+l^-+\cdots$ for $m_H=300$ 
GeV. 
Fig. 3(c) shows the normalized energy correlation for the decay of a 
pseudoscalar 
Higgs $H\to W^+W^-\to l^+l^-+\cdots$ for $m_H=300$ GeV.
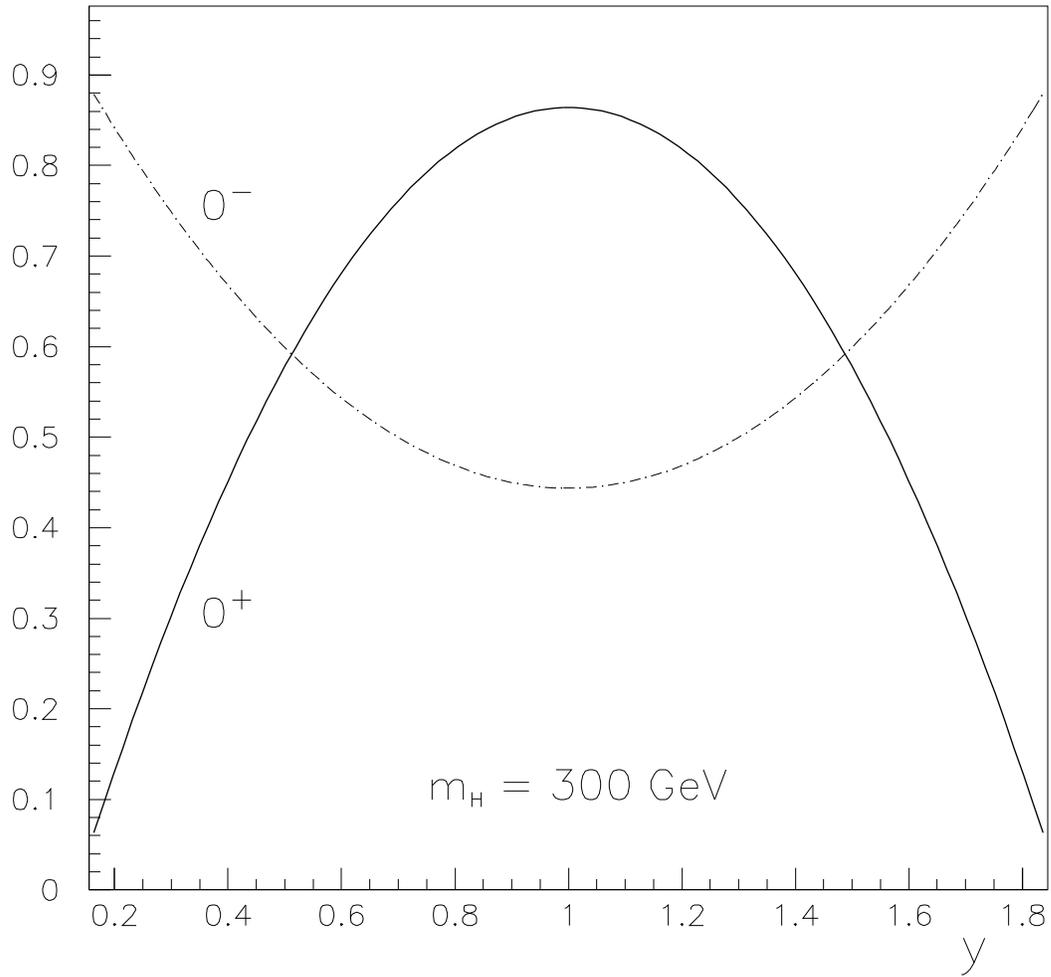
\item[Fig. 4.] Energy distribution of a single lepton in the decay 
$H\to W^+W^-
\to l^{\pm}+\cdots$ for $m_H=300$ GeV. The full curve represents the scalar 
case and 
the dotted curve shows the pseudoscalar case.
\end{itemize}

\end{document}